\documentclass[aps,prb,twocolumn,superscriptaddress,floatfix,showpacs]{revtex4}
\usepackage{psfig}

\bibstyle{apsrev.bib}

\begin{document}

\title{Low-energy fixed points of random Heisenberg models}

\author{Y.-C. Lin}
\affiliation{
Institut f\"ur Physik, WA 331, Johannes Gutenberg-Universit\"at, 
55099 Mainz, Germany}
\author{R. M\'elin}
\affiliation{ 
Centre de Recherches sur les Tr\'es Basses
Temp\'eratures\thanks{U.P.R. 5001 du CNRS, Laboratoire conventionn\'e
avec l'Universit\'e Joseph Fourier}, B. P. 166, F-38042 Grenoble,
France}
\author{Heiko Rieger}
\affiliation{
Theoretische Physik, Universit\"at des Saarlandes, 
66041 Saarbr\"ucken, Germany}
\author{F. Igl\'oi}
\affiliation{
Research Institute for Solid State Physics and Optics, 
H-1525 Budapest, P.O.Box 49, Hungary}
\affiliation{
Institute of Theoretical Physics,
Szeged University, H-6720 Szeged, Hungary}

\begin{abstract}
The effect of quenched disorder on the low-energy and low-temperature
properties of various two- and three-dimensional Heisenberg models is
studied by a numerical strong disorder renormalization group method.
For strong enough disorder we have identified two relevant fixed
points, in which the gap exponent, $\omega$, describing the low-energy
tail of the gap distribution, $P(\Delta) \sim \Delta^{\omega}$ is
independent of disorder, the strength of couplings and the value of
the spin.  The dynamical behavior of non-frustrated random
antiferromagnetic models is controlled by a singlet-like fixed point,
whereas for frustrated models the fixed point corresponds to a large
spin formation and the gap exponent is given by $\omega \approx
0$. Another type of universality classes is observed at quantum
critical points and in dimerized phases but no infinite randomness
behavior is found, in contrast to one-dimensional models.
\end{abstract}


\maketitle

\newcommand{\bc}{\begin{center}}
\newcommand{\ec}{\end{center}}
\newcommand{\be}{\begin{equation}}
\newcommand{\ee}{\end{equation}}
\newcommand{\ba}{\begin{array}}
\newcommand{\ea}{\end{array}}
\newcommand{\beqn}{\begin{eqnarray}}
\newcommand{\eeqn}{\end{eqnarray}}

\section{Introduction}

The Heisenberg model plays a central role in the theory of
magnetic ordering\cite{patrik} and the two-dimensional ($2d$)
antiferromagnetic (AF) model has been intensively studied motivated by
its relation to high-temperature superconductivity\cite{fradkin}.
According to the Mermin-Wagner theorem\cite{merminwagner}, no
long-range order (LRO) can persist at finite temperatures in the
homogeneous Heisenberg model if $d \le 2$. At zero temperature, the LRO of the
classical ground state is reduced by quantum fluctuations. This
effect is particularly strong in (quasi)-$1d$ AF models and gives 
rise to the complete destruction of N\'eel-type LRO. Fluctuations enhanced by 
quenched randomness and frustration can further destabilize LRO, resulting 
in disordered ground states even in higher-dimensional systems.
In various experiments, in which quasi-two-dimensional magnetic materials that
can appropriately be described by the $2d$ HAF were diluted with
static non-magnetic impurities (Mg or Zn in La$_2$CuO$_4$, and Mg in
K$_2$CoF$_4$ or K$_2$MnF$_4$), a disorder-induced transition from
Ne\'el order to a spin liquid was observed: If the impurity
concentration is larger than a critical value the LRO is
destroyed\cite{experiment1,experiment2}.

The behavior of Heisenberg antiferromagnets (HAF-s) in the 
presence of quenched randomness is generally very complex and 
present understanding of this is not complete. Most of the theoretical
results have been obtained for $1d$ models, many of them by a strong
disorder renormalization group (SDRG) method introduced originally
by Ma, Dasgupta and Hu for the random $S=1/2$ AF spin chain\cite{MDH}.
Fisher\cite{fisher} has shown that the SDRG method leads to 
asymptotically exact results in the vicinity of a quantum critical
point, which corresponds to the chain without dimerization.
At the quantum critical point, the ground state can be described by the notion
of a {\it random singlet} (RS) phase, which consists of effective
singlets of pairs of spins that are arbitrarily far from each
other. Fisher's SDRG treatment has been extended to the dimerized
phases that turned out to be equivalent to quantum Griffiths
phases\cite{ijl01}. The SDRG method has also been applied for random
$S=1$\cite{1dS-1-RG} and $S=3/2$\cite{1dS-32-RG} spin chains and for
various random spin ladder models\cite{ladder_paper}. In general, the
Haldane gapped phases stay gapped for weak disorder, while they become gapless
and often form RS phases for strong disorder.

To study the singular properties of the $S=1/2$ Heisenberg model with
mixed ferromagnetic (F) and AF couplings, the SDRG method has to be
modified.  In $1d$, the presence of ferromagnetic couplings leads to the 
formation of large spin clusters in the RG treatment, with an effective 
moment that grows without limits as the energy scale is 
lowered.\cite{westerberg} As a consequence, the ground state properties of 
random Heisenberg chains with mixed AF and F coupling and of those with only 
AF couplings are different. The presence of large effective spins in the 
low-energy limit was also observed for random AF spin ladders with site 
dilution.\cite{ladder_dil}

Not many theoretical investigation of the effect of quenched disorder
in higher-dimensional random HAF-s exist, and those that have been
done are almost exclusively restricted to dilution on the square
lattice. Quantum Monte Carlo studies of the HAF on a diluted square
lattice show that LRO disappears at the classical percolation
point\cite{kthkmt}. While in earlier investigations a novel,
$S$-dependent critical behavior was found\cite{kthkmt}, recent studies
identify the transition as an $S$-independent classical percolation
transition with the well known exponents\cite{sandvik}. Another work
studied the $\pm J$ Heisenberg (quantum) spin glass and found that for
a concentration of F bonds $p>p_c\approx0.11$ the N\'eel-type LRO in
the ground state vanishes and is replaced by a so-called spin glass
phase\cite{2dSG-jap}. Within the spin-glass phase, the average ground
state spin, $S_{\rm tot}$, scales as $S_{\rm tot} \sim \sqrt{N}$, and
the gap as $\Delta E \sim 1/N$, where $N$ is the number of
spins.\cite{2dSG}

In this paper we study the effect of randomness in higher dimensional HAF-s 
by means of the SDRG method. In particular, we consider the low-energy 
behavior of frustrated and non-frustrated systems in $2d$ and $3d$. 
As we will mention in the next section the pure
(i.e. non-random) versions of these models have a ground state that
has either AF or dimer LRO or is disordered, i.e.\ in a spin-liquid
state. By calculating the gap distribution and cluster formation
within the SDRG scheme we shall characterize the change of the pure
ground state structure of the pure systems by the effect of the disorder.

The paper is organized as follows: The models and their phase diagrams
for non-random couplings are presented in Sec. II. The SDRG method and
its different low-energy fixed-points for (quasi)-$1d$ systems are
discussed in Sec. III.  Results of the SDRG method on different $2d$
and $3d$ models are presented in Sec. IV and discussed in Sec. V.

\section{The models and their phase diagram for nonrandom couplings}

We start with the Hamiltonian of a nearest-neighbor spin-1/2 AF
Heisenberg model:
\be
H_1=\sum_{\langle k k' \rangle\;\rm nn} J {\bf S}_k {\bf S}_{k'}\;,
\label{H_1}
\ee
where $J>0$ and the summation runs over nearest neighbor(nn) pairs,
$\langle k,k' \rangle$, of a regular lattice. In $1d$, N\'eel-type
LRO is destroyed by quantum fluctuations and the system with half-integer 
spin value $S$ show quasi-long-range-order (QLRO), i.e. correlations 
in the ground state decay algebraically \cite{lutherpeschel,haldane}. 
A similar behavior can be observed in AF spin ladders with an odd number of
legs\cite{ladder}. Both systems have a gapless excitation spectrum and
in finite chains of length $L$ the gap vanishes algebraically with a
dynamical critical exponent $z_c=1$, which is characteristic for a
quantum critical point (QCP):
\be
\Delta E_{\rm cr}(L) \sim 1/L^{z_c}\;.
\label{E_crit}
\ee
Quantum fluctuations play a different role in AF spin chains with an
integer spin\cite{haldane} and for spin ladders with an even number of
legs\cite{ladder}. These systems show a topological string order,
which is accompanied by exponentially decaying correlations and by a
finite gap in the spectrum.

One can approach a $2d$ geometry by successively increasing the number
of legs of AF spin ladders. The resulting square lattice has a
qualitatively different low-energy behavior: The effect of quantum
fluctuations is weaker and the ground state shows N\'eel-type
LRO\cite{regeryoung}. Compared with the classical ground state the
sub-lattice magnetization for $S=1/2$ is reduced by about $40\%$.
Generally in ordered AF phases the excitation spectrum is gapless. In
finite $d$-dimensional system of linear size $L$ - according to
spin-wave theory and analysis of the non-linear sigma model - the gap
behaves as\cite{neubergerziman}:
\be
\Delta E_{\rm od}(L) \sim 1/L^{d} \;.
\label{E_od}
\ee

Frustration generally leads to a further reduction of the N\'eel
LRO. Frustration of geometrical origin is present in the triangular
lattice, where the sub-lattice magnetization is about $50 \%$ of its
classical value\cite{triangular}. In more loosely packed frustrated
lattices, such as in the kagom\'e lattice\cite{kagome} and in the
square lattice with crosses\cite{2dpyr}(see, however
Ref.~\onlinecite{2dpyr_new}) or in the $3d$ pyrochlore lattice\cite{pyr}, 
the LRO completely disappears and the systems have a
disordered ground state. The correlations are short ranged and one
finds a finite triplet gap in which a continuum of singlet excitations
exists. In the case of the kagom\'e lattice these extend down to the
ground state\cite{kagome}.

\begin{figure}[t]
\centerline{\psfig{file=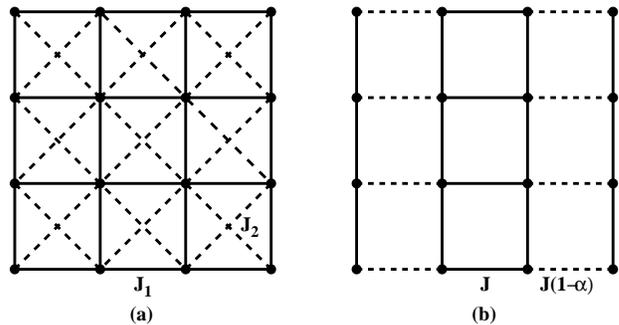,height=4.25cm}}
\vskip 0.2truecm
\caption{ The $J_1-J_2$ model (a) and the dimerized model (b) on the
square lattice.}
\label{FIG01}
\end{figure}

Competing interactions are another source of frustration which
can also lead to disordered ground states. As an example, we consider the AF
$J_1-J_2$ model with first- ($J_1$) and second-neighbor ($J_2$)
interactions, described by the Hamiltonian: 
\be
H=H_1+H_2\;,
\ee
where
\be
H_2=\sum_{\langle k k' \rangle\;\rm nnn} J_2 {\bf S}_k {\bf S}_{k'}\;,
\label{H_nnn}
\ee
and the coupling in $H_1$ (Eq.(\ref{H_1})) is denoted as $J \equiv J_1$
(see Fig.~\ref{FIG01}a). In $2d$, there are at least three phases as shown 
in Fig~2.a. For small frustration, $J_2/J_1=\rho$, the system possesses 
AF LRO, whereas for large frustration the system goes to the collinear 
state, in which ferromagnetically ordered columns of spins are arranged 
antiferromagnetically. In the range $0.34<\rho<0.60$, the ground state is 
disordered and the spectrum is gapped for all types of 
excitations\cite{J1J2}. According to recent numerical studies\cite{SOW} 
there are probably several quantum phases in this region, separated by 
different type of quantum phase transitions. 

\begin{figure}[t]
\centerline{\psfig{file=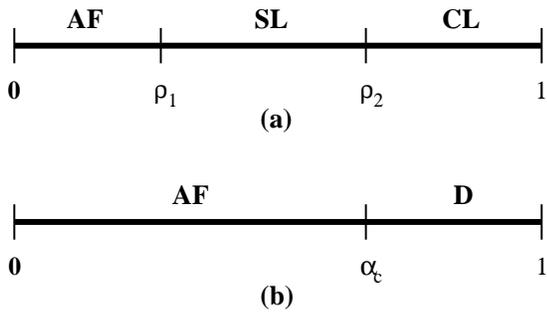,height=4.cm}}
\vskip 0.2truecm
\caption{Phase diagrams of square lattice HAF models. a) For the
$J_1$-$J_2$ model with varying frustration, $\rho=J_2/J_1$, there are
three regions: the ordered AF phase and the ordered collinear (CL)
phase, separated by a disordered spin-liquid (SL) region. b) In the
dimerized model, the AF and dimer (D) ordered phases are separated by a
quantum critical point at $\alpha_c$.}
\label{FIG02}
\end{figure}

Finally, we introduce a dimerization into model (\ref{H_1}). We
consider the square lattice, denote a lattice site $k$ by its two
coordinates, $k=(i,j)$ and define
\be
H_{\rm dim}=-\sum_{i,j} J \alpha {\bf S}_{2i,j} {\bf S}_{2i+1,j}\;.
\label{H_dim}
\ee
The dimerized model is then described by the Hamiltonian $H=H_1+H_{\rm
dim}$ and has a layered structure, see in Fig.~\ref{FIG01}b. Its phase
diagram is shown in Fig.~\ref{FIG02}b as a function of the
dimerization parameter $0<\alpha<1$. For $\alpha<\alpha_c=0.686$ the
ground state has AF LRO, whereas for $\alpha>\alpha_c$ the system is
in an ordered dimerized phase, in which spin-spin correlations along
vertical lines approach different limits if the distance between the
spins is odd or even lattice sites, respectively.  In the dimerized
phase, there is a finite gap which vanishes at $\alpha_c$ as $\Delta E
\sim (\alpha-\alpha_c)^{\nu}$ with $\nu=0.71$, characteristic for the
universality class of the $3d$ classical Heisenberg
model\cite{MYTT-2ddimer,note}. We note that dimerization with another
topology has been studied recently in Ref.~\onlinecite{klumper}.

The random Heisenberg models we investigate in this paper include 
the 2-/3-dimensional AF model on the regular lattice (\ref{H_1}), the 
dimerized AF model (\ref{H_dim}) in $2d$, geometrically frustrated AF models
on the triangular lattice as well as on the kagom\'e lattice, the 2-/3-
dimensional $J_1-J_2$ model and the $3d$ AF-F models. We are
interested in how the phase diagrams in Fig.~\ref{FIG02} will be modified 
due to the presence of strong quenched randomness.

\section{The SDRG method and its low-energy fixed points in $1d$ models}

\begin{figure}[b]
\centerline{\psfig{file=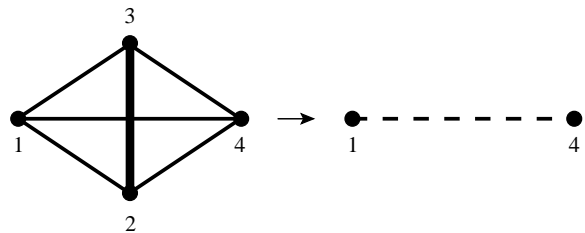,height=3.cm}}
\caption{Singlet formation and decimation for a spin configuration
  that does not have a chain topology and typically occurs in higher
  dimensional systems.}
\label{FIG03}
\end{figure}

The basic ingredient of the SDRG method in Heisenberg models is a
successive decrease of the energy scale of excitations via a
successive decimation of couplings. We start with a $S=1/2$
HAF model in which the strongest
coupling is, say $J_{23}$, the one between lattice sites $2$ and $3$
(c.f. Fig.~\ref{FIG03}). If $J_{23}$ is much larger than its
neighboring couplings $J_{12},J_{13},J_{24}$ and $J_{34}$, the spins at $2$
and $3$ form an effective singlet and are decimated. The effective
coupling between the remaining sites, $1$ and $4$ in second order
perturbation theory is given by:
\be
{\tilde J}^{\rm eff}_{14}=\lambda 
\frac{(J_{12}-J_{13})(J_{34}-J_{24})}{J_{23}},\quad
\lambda(S=1/2)=1/2\;.
\label{deci_sing}
\ee
In a chain geometry the couplings $J_{13}$ and $J_{24}$ would not be
present and the resulting RG flow always generates AF couplings.  
However, for extended, not strictly $1d$ objects, some of the
generated new couplings can be ferromagnetic (e.g. if $J_{12}<J_{13}$
and $J_{34}>J_{24}$ or vice versa) and therefore the decimation rules
have to be extended. If at one RG step an F bond turns out to be the
strongest one, its decimation will lead to an effective spin
$\tilde{S}=1$. In the following steps, the system will renormalize to a set
of effective spins of different magnitude interacting via F {\it
and/or} AF couplings.

For higher dimensional systems, the basic decimation processes are  
the singlet formation in Eq.(\ref{deci_sing}) and 
the effective spin (cluster) formation. To specify the latter, let us 
consider three spins $S_1,S_2$ and $S_3$ with interactions fulfilling 
$|J_{23}| \gg |J_{12}|,|J_{13}|$.  In the action of the RG,  the two 
original spins $S_2$ and $S_3$ form a new effective spin of magnitude 
$\tilde{S}=|S_2 \pm S_3|$ representing the total spin of 
the ground state in the two-spin Hamiltonian 
$H_{23}=J_{23}{\bf S}_2 {\bf S}_2$, 
where the positive (negative) sign refers to an F (AF) coupling.
The corresponding energy gap, $\Delta$, between the ground state and the
first excited state in the Hamiltonian $H_{23}$ is given by 
$\Delta=|J_{23}|(S_2+S_3)$ and $\Delta=J_{23}(|S_2-S_3|+1)$, 
for an F and AF coupling, respectively. If $J_{23} >0$ (AF) and $S_2=S_3$, 
it follows an effective singlet formation as described above. 
If $\tilde{S} \neq 0$, within first order perturbation theory 
the new coupling between $S_1$ and $\tilde{S}_{23}$ is given by
\be
\tilde{J}^{\rm eff}=c_{12}J_{12}+c_{13}J_{13}\;,
\label{deci_cl}
\ee
with 
\[ c_{12}=\frac{\tilde{S}(\tilde{S}+1)+S_2(S_2+1)-S_3(S_3+1)}
               {2\tilde{S}(\tilde{S}+1)} \]
and
\[ c_{13}=\frac{\tilde{S}(\tilde{S}+1)+S_3(S_3+1)-S_2(S_2+1)}
               {2\tilde{S}(\tilde{S}+1)}. \]
At each RG step, we find the pair of the spins with the largest
energy gap $\Delta$ that sets the energy scale, $\Omega$, and decimate 
them according to renormalization rules described in (\ref{deci_sing}) or
(\ref{deci_cl}). A detailed derivation of these
renormalization rules can be found in Ref.~\onlinecite{Melin-RG}.

The fixed point of the RG transformation for lattices that do {\it
not} have a chain geometry may depend on their topology, the original
distribution of bonds, the strength of the disorder, etc. We briefly
summarize the existing results for spin chains and ladders since it
might be helpful for analyzing the RG results in higher dimensional
systems.

In the case of the random AF chain (which does neither have F bonds nor
frustration), the RG procedure described above runs into an infinite
randomness fixed point (IRFP) corresponding to a random singlet
phase. In this phase the renormalized clusters are singlets,
thus the total magnetic moment is zero, and the energy and length scales are
related via
\be
-\ln \Omega \sim L^{1/2}\;,
\label{IRFP}
\ee
which means that the dynamical exponent is formally infinite.

A dimerized $S=1/2$ chain with random AF even ($J_{\rm e}$) and odd
($J_{\rm o}$) couplings shows dimer order and the low-energy behavior
is controlled by a {\it random dimer fixed point} at which the
dynamical exponent, $z$, is finite and a continuously varying function
of the strength of the dimerization measured by $\delta_{\rm dim}=[\ln
J_{\rm e}]_{\rm av}-[\ln J_{\rm o}]_{\rm av}$\cite{ijl01,ijr00}. At
this fixed point, the low-energy-tail of the distribution of the
effective couplings, $J_e$, is given by:
\be 
P(J_{\rm e},\Omega){\rm d}J_{\rm e} \simeq \frac{1}{z}
\left(\frac{J_{\rm e}}{\Omega}\right)^{-1+1/z} 
\frac{{\rm d}J_{\rm e}}{\Omega}\;,
\label{Jz}
\ee
for $\delta_{\rm dim}>0$. This random dimer phase is a Griffiths
phase\cite{griffiths} and we refer to it as a Griffiths fixed point
(GFP). At this GFP, the gap of finite chains of length $L$ obey a
distribution similar to Eq.(\ref{Jz}):
\be
P_L(\Delta)=L^z \tilde{P}(L^z \Delta) 
\sim L^{z(1+\omega)} \Delta^{\omega}\;,
\label{PDelta}
\ee
which is characterized by the gap exponent, $\omega$. As a consequence
of Eq.(\ref{PDelta}), which holds in any dimension, several dynamical
quantities at a GFP are singular and the characteristic exponents can
all expressed via $\omega$.  For example the susceptibility $\chi$,
the specific heat $C_v$ (at a small temperatures $T$), and the
magnetization $m$ (in a small field $h$), behave as:
\be
\chi(T) \sim T^{-\omega},\quad C_v(T)\sim T^{\omega+1},\quad m(h) 
\sim h^{\omega+1}\;.
\label{Griffiths}
\ee
In the Griffiths phase there is a simple relation between the
dynamical exponent, $z$, and the gap exponent, $\omega$, which can be
obtained by the following phenomenological consideration\cite{phenom}:
If the Griffiths singularities are due to rare events (produced by the
couplings) that give rise to {\it localized} low-energy excitations,
the gap distribution should be proportional to the volume,
$P_L(\Delta) \sim L^d$. From Eq.(\ref{PDelta}) then follows:
\be
z=\frac{d}{1+\omega}\;,
\label{z_omega}
\ee
which is consistent with the exact result in the random dimer phase in
Eq.(\ref{Jz}). However, if the low-energy excitations are {\it extended}
the relation (\ref{z_omega}) might not hold.

In a spin chain with mixed F and AF couplings\cite{westerberg}, large
effective spins, $S_{\rm eff}$, are formed at the fixed point of the
transformation. The size of of these spin clusters scales with the
fraction of surviving sites during decimation, $1/N$, as:
\be
S_{\rm eff} \sim N^{\zeta}\;.
\label{S_N}
\ee
The following random walk argument \cite{westerberg} gives $\zeta=1/2$:
The total moment of a typical cluster of size $N$ can be expressed as
$S_{\rm eff}=|\sum_{1=1}^N \pm S_i|$, where neighboring spins with F
(AF) couplings enter the sum with the same (different) sign. If the
position of the F and AF bonds are uncorrelated and if their
distribution is symmetrical, one has $S_{\rm eff}\propto N^{1/2}$,
i.e.\ Eq.(\ref{S_N}) with $\zeta=1/2$.

A non-trivial relation constitutes the connection between the energy
scale $\Omega$ and the size of the effective spin:
\be
S_{\rm eff} \sim \Omega^{-\kappa}\;,
\label{S_Omega}
\ee
where a numerical estimate of the exponent is 
$\kappa=0.22(1)$\cite{westerberg}. Comparing Eq.(\ref{S_N}) with
Eq.(\ref{S_Omega}), the relation between the length scale 
$L\sim N^{1/d}$ ($d=1$) and the energy scale is:
\be 
\Omega \sim L^{-z},\quad z=\frac{d \zeta}{\kappa}=\frac{1}{2 \kappa}\;,
\label{L_Omega}
\ee
where $z$ is the dynamical exponent. The distribution of low-energy
gaps, $P_L(\Delta)$, has the same power-law form as in
Eq.(\ref{PDelta}). Therefore from the scaling behavior of
$P_L(\Delta)$ the gap exponent, $\omega$, and the dynamical exponent,
$z$, can be obtained.  Due to the large moment formation the singularities
of the dynamical quantities are different from those 
in the random dimer phase in Eq.(\ref{Griffiths}), i.e. at a GFP. 
Generalizing the reasoning in Ref.~\onlinecite{westerberg}, 
we obtain in $d$-dimensions:
\be
\chi(T) \sim T^{-1},~ C_v(T)\sim T^{2\zeta(\omega+1) }|\ln T|,
~m(h) \sim h^{\frac{\zeta(1+\omega)}{1+\zeta(1+\omega)}}\;,
\label{F_AF}
\ee
thus the singularities involve both exponents $\zeta$ and $\omega$.
In the following, we refer to this type of fixed point as large spin
fixed point (LSFP).

AF spin ladders, although being quasi-one-dimensional, have a
non-trivial, non-chain-like topology and during renormalization also F
bonds can be generated according to Eq.(\ref{deci_sing}). Different
random AF two-leg ladders were studied in
Ref.~\onlinecite{ladder_paper} with tho following results: If the
disorder is strong enough the gapped phases of the non-random systems
become gapless. The low energy behavior is generally controlled by a
GFP, where the dynamical exponent is finite and depends on the
strength of the disorder. However, at random quantum critical points,
separating phases with different topological or dimer order, the
low-energy behavior is controlled by an IRFP. In diluted AF spin
ladders also LSFP-s have been identified\cite{ladder_dil}.

To close this section we summarize that in one-dimensional and in
quasi-one-dimensional random Heisenberg systems there are two
different types of low-energy fixed points, which are expected to be
present in higher dimensional systems, too.  Both for a GFP and for a
LSFP, the low-energy excitations follow the same power-law form in
Eq.(\ref{PDelta}) from which the exponents, $\omega$ and $z$ can be
deduced. At a GFP these two exponents are expected to be related
through $z=\frac{d}{1+\omega}$ (\ref{z_omega}). On the other hand, for
a LSFP, where the excitations are not localized, this relation
probably does not hold. At such a LSFP there is a third independent
exponent, $\zeta$ involved in the dynamical singularities
partially listed in Eq.(\ref{F_AF}).

In the next section we are going to study different two- and
three-dimensional random Heisenberg models. In particular we will be
interested in the possible difference in the low-energy fixed point
for non-frustrated and frustrated systems. Since extended
(quasi-one-dimensional or higher-dimensional) random HAF models and
Heisenberg models with mixed F and AF bonds follow the same
renormalization route, they could, in principle, be attracted by the
same fixed points, but also new fixed points can emerge, as we will
see.

\section{Renormalization of higher-dimensional systems}

This section is the central part of our work, where we present our
results for the ground state structure of various two- and
three-dimensional random Heisenberg models obtained by the numerical
application of the SDRG. In practice we start with a finite system of
linear size $L$ with periodic boundary conditions and perform the
decimation procedure up to the last effective spin (or decimate out
the last spin-singlet). The energy scale corresponding to the last
decimation step is denoted by $\Delta$. This procedure is performed
for several thousand realizations of the disorder and yields a
histogram for $\Delta$, which represents our estimate of the
probability distribution $P_L(\Delta)$. From this we extract the gap
exponent $\omega$ and the dynamical exponent $z$ via the asymptotic
relation given in Eq.(\ref{PDelta}). Moreover, from the average size
of the effective spin at the last step, $\mu_L=[S_{\rm eff}]_{\rm
av}$, the cluster exponent, $\zeta$, in Eq.({\ref{S_N}) is
deduced. The value of $\omega$, $z$ and $\zeta$ will then be used to
discriminate the different possible low-energy fixed points described
in the previous section.

Throughout this paper we use a power-law distribution for the random
couplings $0<J \le 1$ for AF models:
\be
P_D(J) =\frac{1}{D} J^{-1+1/D} \;,
\label{P_D}
\ee
where $D^2=[(\ln J)^2]_{\rm av}-[\ln J]_{\rm av}^2$ denotes the
strength of the disorder. Note that both the initial distribution of
the couplings in Eq.(\ref{P_D}) and the final distribution of gaps in
Eq.(\ref{PDelta}) follow power laws. If $1/(\omega+1)<D$, the strength 
of disorder is reduced during renormalization, thus the
low-energy random fixed point is a conventional one. More generally,
for a conventional random fixed point $\omega>-1$.  
In contrast to this, at a IRFP the disorder growths without limits, 
thus here formally $\omega=-1$ and the dynamical exponent is infinite. 
We often use the uniform distribution, which corresponds to $D=1$ 
in Eq.(\ref{P_D}). For models with random $F$ and $AF$ couplings we take 
either a Gaussian distribution:
\be
P_G(J) =\frac{1}{\sqrt{2 \pi \sigma^2}} \exp(-J^2/2\sigma^2) \;,
\label{P_G}
\ee
or a rectangular distribution:
\be
P_r(J) =\Theta(J-r+1/2) \Theta(r+1/2-J)\;,
\label{P_r}
\ee
where $\Theta(x)=1$, for $x>0$ and zero, otherwise. The latter distribution
is symmetric for $r=0$, whereas for $r=1/2$ we recover the uniform distribution
of AF couplings in Eq.(\ref{P_D}) with $D=1$.

\subsection{Two-dimensional models}

In the calculations for $2d$ we usually considered systems of linear
size up to $L=32$, but for some cases in which the convergence was
faster we went only up to $L=10-16$. The typical number of
realizations were several hundred-thousands for the smaller sizes and
several ten-thousands for larger systems for each value of $D$. At the
first part we investigate non-frustrated models, such as the HAF on
the square lattice with and without dimerization. In the second part
of our study we consider frustration, the origin of which could
be i) geometrical as for instance for the triangular and kagom\'e
lattice ii) due to a random mixture of F and AF couplings as for
instance for the $\pm J$ spin glass model and iii) due to competition
between first- and second-neighbor couplings as for the $J_1$-$J_2$ model.

\subsubsection{HAF on the square lattice}

We start with the renormalization of the HAF on the square lattice.
The probability distribution of the gap calculated for a uniform bond
distribution (Eq.(\ref{P_D}) with $D=1$) is shown in Fig.~\ref{FIG04}
for different linear sizes. In a log-log plot the small gap region of
the curve is linear, the slope of which, according to
Eq.(\ref{PDelta}), corresponds to $\omega+1$.  With increasing size
one observes a slight broadening of the distributions indicating a decreasing
effective gap exponent which, however, seems to converge to a finite
asymptotic value:
\be
\omega_{AF} = 0.7(1),\quad d=2\;.
\label{omega_AF}
\ee

\begin{figure}[b]
\centerline{\psfig{file=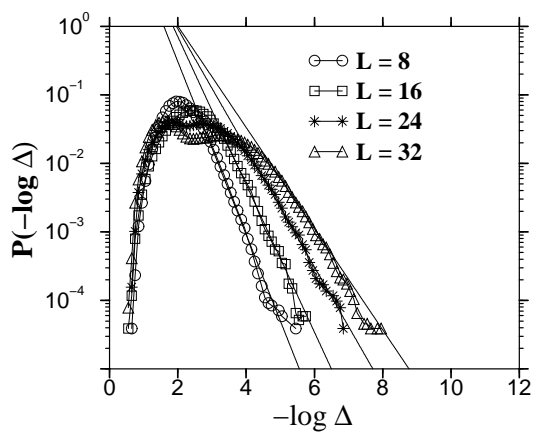,height=5.75cm}}
\vskip 0.2truecm
\caption{ Distribution of the energy gap of the square-lattice HAF
  with uniformly distributed random couplings, for linear sizes
  $L=8,16,24$ and $32$. The slope of the low-energy tail of the
  distributions is given by $-(\omega+1)=-d/z$. The straight line
  for $L=32$ has a slope $\approx -1.7$.}
\label{FIG04}
\end{figure}

During renormalization we observed simultaneously an effective singlet
formation, thus in Eq.(\ref{S_N}) one has $\zeta=0$. Our estimate for
the dynamical exponent satisfies the relation in Eq.(\ref{z_omega}),
yielding $z_{AF}=1.2$. Thus we conclude that the low-energy fixed point of
the system is a {\it conventional, finite disorder} Griffiths fixed
point and the thermodynamical singularities are given by
Eq.(\ref{Griffiths}). For other disorder strengths $D$ we reach the
same conclusions and our estimates for the gap exponents for each $D$
agree with the value in Eq.(\ref{omega_AF}) within the error
bars. Thus the low-energy singular behavior of the 2d random HAF does
not depend on the strength of disorder, in contrast to random quantum
spin ladders\cite{ladder_paper}.

\subsubsection{Square-lattice HAF with dimerization}

Next we study the low-energy behavior of the dimerized HAF, as
sketched in Fig.~\ref{FIG01}b. For site and bond dilution the
stability of the gapped, dimerized phase was recently
investigated\cite{YTMT02}.  Here we consider the effect of strong AF
bond disorder. In our calculation we used uniform initial
randomness and performed the renormalization for several values of the
dimerization parameter, $\alpha$. The possible values of the two types
of couplings were in the regions $(0,1]$ and $(0,(1-\alpha)]$,
respectively. For any value of $\alpha$ in the range $0<\alpha<1$, 
we observed an effective singlet formation and the estimated gap exponents 
$\omega$ and dynamical exponents $z$ are found to satisfy the relation in 
Eq.(\ref{z_omega}). The extrapolated
dynamical exponents as plotted in Fig.~\ref{FIG05} seem to be
approximately constant in two regions, which corresponds to the two
phases of the pure model in Fig.~\ref{FIG02}b.  For weaker
dimerization the dynamical exponent corresponds to the one of the
random HAF, and for stronger dimerization $z$ is approximately
equal to the one of the disconnected two-leg ladder system, to which
the case $\alpha=-1$ reduces, with $z \approx
1.07$\cite{ladder_paper}. We expect that the dimer order is finite in
the RD region, whereas it is zero (or very small) in the random HAF
region.  Between the two regions, corresponding to the neighborhood of
the phase transition point in the pure system in Fig.~\ref{FIG02}b,
the dynamical exponent drops to a minimal value. This cross-over could
happen in a smooth, non-singular way, or in a sharp phase
transition separating the random AF and the random dimer phases.  Due
to strong finite-size effects we could not discriminate the two
scenarios.

\begin{figure}[t]
\centerline{\psfig{file=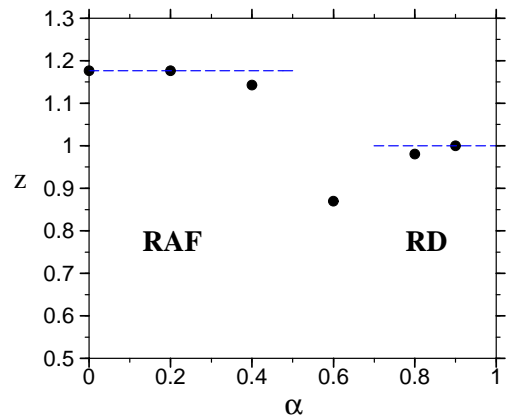,height=5.75cm}}
\vskip 0.2truecm
\caption{Extrapolated dynamical exponent of the random dimerized HAF
 on the square lattice. The random AF and the random dimerized phases
 are separated by a cross-over region in which the dynamical
 exponent is minimal.}
\label{FIG05}
\end{figure}

We note that $z$ in the cross-over region behaves in the opposite way
as in the dimerized ladders, where the dynamical exponent at the
transition point in a finite system is maximal, and 
increases without limits\cite{ladder_paper} for increasing system
size, signaling an IRFP. In the two-dimensional case considered here
the combined effect of critical fluctuations and quenched randomness
seem to reduce the value of the dynamical exponent. 
Our calculations indicate that in the random dimer phase the low-energy 
behavior is controlled by a GFP and the dynamical singularities are given by
Eq.(\ref{Griffiths}).

\subsubsection{Randomly frustrated models ($2d$)}

\begin{figure}[t]
\centerline{\psfig{file=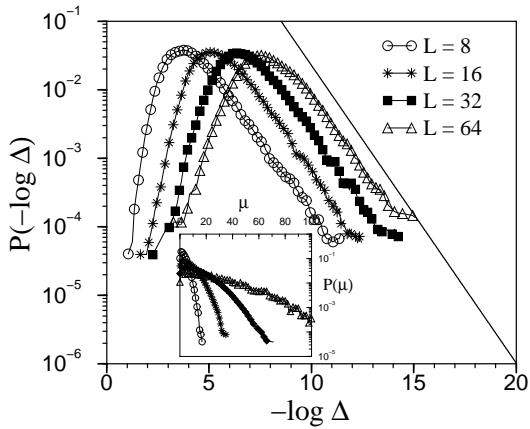,height=5.75cm}}
\vskip 0.2truecm
\caption{ Probability distribution of the energy gap on the square
  lattice with mixed F and AF bonds following a Gaussian distribution
  with $\sigma=1$. (The slope of the straight line is $-1$).  Inset:
  Distribution of the spin moments.}
\label{FIG06}
\end{figure}

In this subsection we consider the Heisenberg model on the square
lattice with a random mixture of F and AF couplings. This is a model
for a {\it quantum spin glass} \cite{2dSG-jap,2dSG} and we denote the
corresponding fixed point as {\it spin glass fixed point} (SGFP),
although we do not explicitely check for the existence of proper spin
glass order in the ground state (for instance via the calculation of
the Edwards-Anderson spin glass order parameter\cite{sg-review}). As
we will see, this fixed point differs from the other fixed points we
found for non-frustrated models, so we feel that the use of this
notation is justified. In particular we find a large spin formation
proportional to $L$ during RG procedure implying a ground state spin
$S\propto\sqrt{N}$, which is reminiscent of the spin glass behavior
found in \cite{2dSG-jap,2dSG} for this model with alternative methods.

First we report the results for the Gaussian randomness in
Eq.(\ref{P_G}). For this case the distributions of the gaps and of the
effective spin moments are shown in Fig.~\ref{FIG06}.
The gap-distributions for different finite sizes have a very similar
structure: they are merely shifted to each other by a constant
proportional to $\ln L$.  The slope of the low-energy tail of the
distributions is practically independent of the strength of disorder
and in all cases the gap exponent is equal to:
\be
\omega_{SG}=0,\quad d=2 \;,
\label{omega_SG}
\ee
within an accuracy of a few percent.  From the finite-size scaling
properties of the gap distribution, we infer that the relation in
Eq.(\ref{z_omega}) is satisfied and therfore the excitations are
localized, implying 
\be
z_{SG}=2,\quad d=2 \;,
\label{z_SG}
\ee
within an accuracy of a few percent. 
 
On the other hand, the distribution of the effective spin moments in
the inset to Fig.~\ref{FIG06} shows a tendency to broaden with
increasing system size and its average value has a linear $L$
dependence, $[\mu_L]_{\rm av} \approx .42 L$. Therefore the moment
exponent in Eq.(\ref{S_N}) is
\be
\zeta_{SG}=1/2, \quad d=2\;,
\label{zeta_SG}
\ee

We have repeated the above analysis using the symmetric rectangular
distribution in Eq.(\ref{P_r}) both for the $S=1/2$ and the $S=1$
models and we obtained the same critical exponents as in the Gaussian
case. Thus we can conclude that the low-energy behavior in randomly
frustrated $2d$ models is controlled by the same SGFP, independent
of the type of randomness and the size of the spin.

\subsubsection{Geometrically frustrated models}

In this section we consider the HAF on two geometrically frustrated
lattices that have qualitatively different ground states in the
non-random case. The triangular lattice has finite AF long-range order
and low-energy excitations behave as 
in Eq.(\ref{E_od}). In contrast to this, the ground state of the 
kagom\'e lattice is disordered and the low-energy singlet excitations 
have a more complicated size dependence.

\begin{figure}[b]
\centerline{\psfig{file=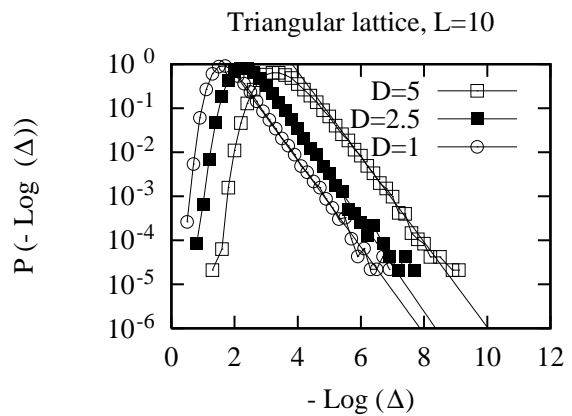,height=5.75cm}}
\vskip 0.2truecm
\caption{ Probability distribution of the energy gap for the
triangular lattice HAF for different strength of randomness in
Eq.(\ref{P_D}).  The low energy tail of the distributions, which has
practically no finite-size dependence for $L \ge 10$, is consistent
with the same gap exponent, $\omega=0$, implying a dynamical exponent
$z_{SG}=d=2$.}
\label{FIG07}
\end{figure}

We start with the HAF on the triangular lattice using the power-law
distribution in Eq.(\ref{P_D}) for the random couplings. The
distribution function of the gap is presented in Fig.~\ref{FIG07} for
different disorder strengths. The slope of the low-energy tail of the
distributions is again, as for the randomly frustrated model of the
last susection, practically independent of the strength of disorder
and in all cases the gap exponent is equal to $\omega=0$
within an accuracy of a few percent.

When calculating the moment of the spin clusters, we notice large spin
formation during the action of the RG. From the size dependence of the
average moment we obtain the exponent in Eq.(\ref{S_N}) $\zeta=1/2$, 
independent of the strength of disorder. From the finite-size scaling
properties of the gap distribution, we infer that the relation in
Eq.(\ref{z_omega}) is satisfied and therfore the excitations are
localized, implying $z_{SG}=d=2$. Thus we can conclude that the
thermodynamical quantities in the random triangular HAF obey the
relations in Eq.(\ref{F_AF}).

Next we focus on the kagom\'e lattice and enlarge the parameter space
by considering the dimerized model, as introduced in
Ref.~\onlinecite{mila_kagome}: Couplings in up-pointing triangles ($J$)
are different from those in down-pointing triangles ($J'$). Analyzing
the results of the RG calculation as already described for the
triangular lattice, we obtain a set of gap, dynamical and moment
exponents for different dimerization, $0.1<J'/J<1.5$, and disorder
strengths, $D=1,~2.5,$ and $5$.  In Fig.~\ref{FIG08} we show our
estimates for the dynamical exponents for $D=2.5$, which
are consistent with the SGFP result in Eq.(\ref{omega_SG}). Also for
other disorder strengths we find the same behavior and we conclude
that the low energy physics of the random kagom\'e HAF is controlled
by the SGFP and the thermodynamic singularities are described by
Eq.(\ref{F_AF}).

\begin{figure}
\centerline{\psfig{file=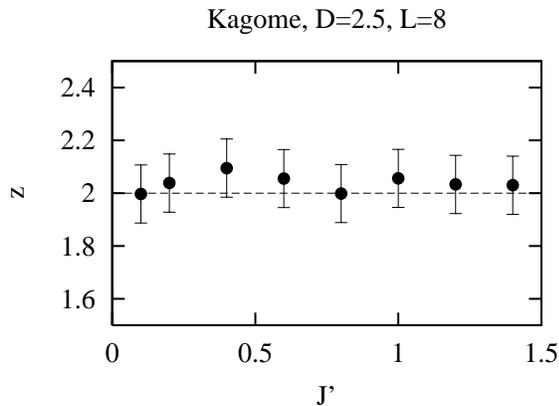,height=5.75cm}}
\vskip 0.2truecm
\caption{Dynamical exponent of the random HAF on the dimerized
  kagom\'e lattice with a randomness parameter $D=2.5$ calculated in
  finite systems having $L^2$ triangles, thus $3 L^2$ sites.}
\label{FIG08}
\end{figure}

\subsubsection{The $J_1$-$J_2$ model}

In our final example for the $2d$ case, the source of frustration 
is the competition between first - $J_1$ - and second-neighbor 
- $J_2$ - couplings, which obey a power-law distribution in Eq.(\ref{P_D}) 
within the range of $0<J_1 \le J_1^{\rm max}$ and $0<J_2 \le J_2^{\rm max}$,
respectively. We have performed the previous analysis at different
points of the phase diagram, $J_2^{\rm max}/J_1^{\rm max}$, and for
different strength of disorder, $D$. In all cases we found that the
relation in Eq.(\ref{z_omega}) is valid. As an illustration we show in
Fig.~\ref{FIG09} our estimates for the dynamical exponents for a
disorder strength $D=5/3$, which are consistent with the SGFP value in
Eq.(\ref{omega_SG}) in a wide range of $0.2 < J_2^{\rm
max}/J_1^{\rm max} < 2.0 $. The same conclusion holds for other
disorder strengths in the range of $1 \le D \le 5$. During
renormalization there is large spin formation and the calculated
cluster exponent is consistent with $\zeta_{SG}=1/2$.  Thus we can
conclude that in the $J_1$-$J_2$ model the different phases in the pure
model (AF and CL ordered, disordered SL) are washed out by strong
disorder, and the whole frustrated region, $J_2/J_1 >0$, is controlled
by the SGFP.

\begin{figure}[t]
\centerline{\psfig{file=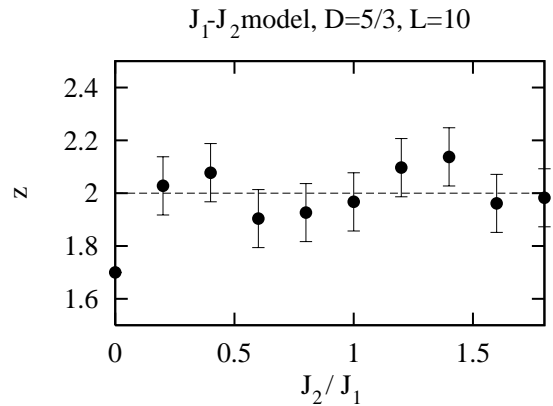,height=5.75cm}}
\vskip 0.2truecm
\caption{Dynamical exponent of the $J_1$-$J_2$ model on the
square lattice with a power-law randomness with $D=5/3$.}
\label{FIG09}
\end{figure}

\subsection{Three-dimensional models}

For the calculations in $3d$ that we present now we considered only
systems of linear size $L=6,8,10$ and $12$, in some cases we went up
to $L=16$. Larger system sizes were computationally not feasible. The
typical number of realizations were several ten-thousand for each
point. Due to the smaller system sizes the finite-size effects in $3d$
are stronger than in $2d$. These finite size effects turned out to be
too strong in the random HAF on the cubic lattice for a safe estimate
for the gap exponent. We can, however, conclude that there is no
large spin formation and the low-energy behavior is controlled by a
conventional GFP.

\subsubsection{Randomly frustrated models ($3d$)}

\begin{figure}[t]
\centerline{\psfig{file=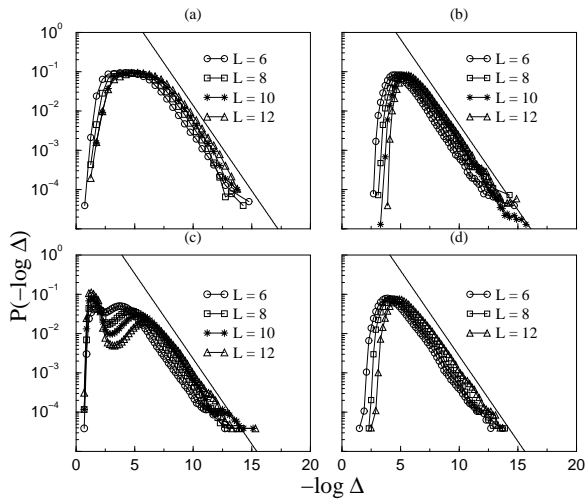,height=6.75cm}}
\vskip 0.2truecm
\caption{Probability distribution of the energy gap on the cubic
 lattice with mixed F and AF bonds. (a) Gaussian distribution,
 $\sigma=1$; (b) symmetric rectangular distribution ($r=0$); (c)
 asymmetric rectangular distribution ($r=0.25$); (d) $S=1$ symmetric
 rectangular distribution.  The low-energy tails of the gap
 distributions for all cases, indicated by straight lines, have a
 slope -1, corresponding to $\omega=0$.}
\label{FIG10}
\end{figure}

We have studied models with mixed F and AF couplings for different
form of initial randomness (Gaussian, symmetric and asymmetric rectangular)
and for comparison calculations on the $S=1$ model are also performed. 
The calculated distributions of the gaps are presented in Fig.~\ref{FIG10}.

As seen in Fig.~\ref{FIG10} the slopes of the low-energy tail of the
gap-distributions are approximately constant, and for our finite
systems they are consistent with a vanishing gap exponent:
\be
\omega \approx 0\quad (d=3)\;.
\label{omega_SG3}
\ee
%
During renormalization, as in $2d$, there is a large spin formation
and the corresponding moment exponent is $\zeta=0.55$, for symmetric
distributions (Gaussian and rectangular) and $\zeta=0.58$ for the
asymmetric rectangular distribution. Thus $\zeta$ appears to be close
to $1/2$ in both cases. We have also studied the scaling behavior of
the reduced gap distribution, $\tilde{P}(L^z
\Delta)=L^{-z}P_L(\Delta)$. In Fig.\ref{FIG11} we show a scaling
collapse of the distributions, which is obtained by $z \approx 1.5$
independently of the disorder distribution. The scaling curves seem to
tend to a finite limiting value at $\Delta=0$, implying a gap exponent
$\omega \approx 0$. We can thus conclude that --- within the range of
validity of the SDRG method --- the relation in Eq.(\ref{z_omega}) is
not valid for frustrated 3d models.

\begin{figure}[t]
\centerline{\psfig{file=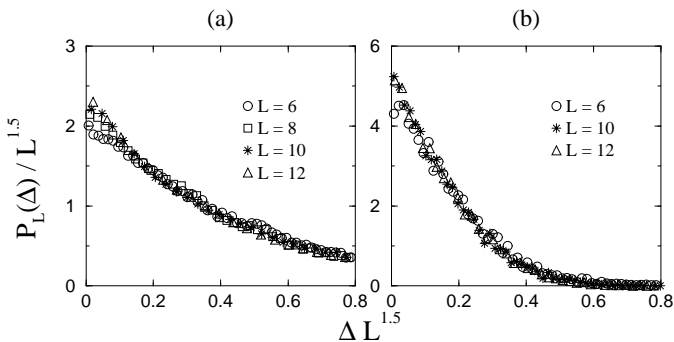,height=4.75cm}}
\vskip 0.2truecm
\caption{Scaling of the reduced gap distribution, $\tilde{P}(L^z
\Delta)=L^{-z}P_L(\Delta)$, for randomly frustrated 3d systems: a)
Gaussian randomness, $\sigma=1$, b) uniform randomness. In both cases
it is $z=1.5$.}
\label{FIG11}
\end{figure}

\subsubsection{The $J_1$-$J_2$ model}

We also considered frustration caused by a competition between nearest
and next-nearest neighbor couplings in order to see to what extent the
universality of the SG phase, observed in $2d$ models, is valid in
$3d$.  Here we study systems at different points of the phase diagram,
$J_2^{\rm max}/J_1^{\rm max}$ and for different initial disorder $D$,
using the same notations as for $2d$. Typical gap distribution are
shown in Fig.~\ref{FIG12}, where we observe that the low-energy tail
of the distributions in each case has approximately the same slope
close to $-1$, which results in a gap exponent, $\omega \approx 0$.
This result is consistent with Eq.(\ref{omega_SG3}) obtained for
randomly frustrated models. During renormalization large spin
formation is observed, and the moment exponent, $\zeta$, is found to
depend on the position in the phase diagram: for $J_2^{\rm
max}/J_1^{\rm max}=0.5$ and $1.0$ it is given by $\zeta=.58$ and
$.78$, respectively ($D=1$ is in both cases). The dynamical exponent
in these cases was about the same ($z \approx 3/2$) as for randomly
frustrated models.

\begin{figure}[h]
\centerline{\psfig{file=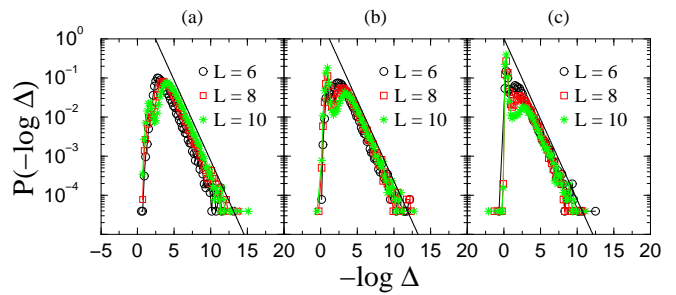,width=\columnwidth}}
\vskip 0.2truecm
\caption{Probability distribution of the energy gap of the $J_1$-$J_2$
model. (a) $J_2^{\rm max}/J_1^{\rm max}=0.5$, $D=1$; (b) $J_2^{\rm
max}/J_1^{\rm max}=1$, $D=2$; (c) $J_2^{\rm max}/J_1^{\rm max}=1$,
$D=1$. (The slope of the straight lines in all cases is
$-1$.)}
\label{FIG12}
\end{figure}

Thus we can conclude that also in $3d$ the low-energy fixed points of
random Heisenberg systems with different types of frustration are
controlled by the same type of SGFP, having the same gap exponent,
$\omega \approx 0$, as in a two-dimensional SGFP. Therefore we
conjecture that the ground states of these $3d$ frustrated models are
in a spin glass phase, too. At these SGFP-s the dynamical
exponent is constant, however the moment exponent has a system
dependence. Thus the low-energy excitations have a universal scaling
behavior, but the thermodynamical singularities in Eq.(\ref{F_AF}),
which depend on the value of $\zeta$, are system dependent.

\section{Discussion}

In this paper we considered higher dimensional HAF-s and studied the
effect of strong randomness on their low-energy/low-temperature
properties by a numerical application of the SDRG method.  Comparing
with the known, partially exact results for $1d$ HAF-s we noticed
several important differences. First, in higher dimensions one
observes a strong universality scenario: there are only a few relevant
fixed points (most important are the random AF fixed point and the
SGFP) and their properties do not depend on the coordination number,
the strength of disorder, value of the spin, etc., but just on the
dimension of the model and the degree of frustration in the system. In
contrast to this, in random spin chains and ladders one has usually a
continuum of low-energy fixed points parameterized by the value of the
dynamical exponent $z$ and which do depend on the aforementioned
details. Second, in higher dimensional HAF-s the singularities are
controlled by (a few) conventional random fixed points, at which the
dynamical exponent is finite. In higher dimensional systems there are
no IRFP-s that can generally be found in (quasi) $1d$ systems at
random quantum critical points. A third difference between $1d$ and
higher-dimensional AF is the following: In $1d$ the renormalization of
random spin-1/2 AF spin chains and random quantum ferromagnets, such
as the random transverse Ising model \cite{fisher,fisheri,ijl01}, lead to
similar IRFPs at quantum critical points. In higher dimensional random
transverse Ising models the random quantum critical point is still an
IRFP\cite{2drg,lkir}, whereas for the random HAF, even at random
quantum critical points, we found in this work the dynamical exponent
to be finite.

One remarkable aspect of our results is the observed universality of
the fixed point controlling the low-energy characteristics of random
frustrated systems \cite{note2}. The gap exponent of this so-called
spin glass fixed point (SGFP) is numerically very close to
zero\cite{boseglass} and we can explain this observation in the
following way.  During renormalization there is a large spin formation
in these systems and therefore we expect that the low-energy excitations in $d
\ge 2$ are extended over the whole (finite) volume of the system (in
a $1d$ topology these excitations are not extended since
unfavorable domains usually restrict the size of excitations). As a
consequence these excitations can be considered as compact objects so
that their reduced (scale invariant) probability density
$\tilde{P}(L^z \Delta)=L^{-z} P_L(\Delta)$ has no size dependence for
a fixed small gap, $\Delta$. This last statement is consistent with a 
vanishing gap exponent, $\omega=0$, according to Eq.(\ref{PDelta}) and
is supported by the numerical results in Fig.\ref{FIG11}.

Finally we make a remark about the accuracy of the results obtained by
the SDRG method. It is generally expected that the SDRG method leads
to asymptotically exact relations concerning singularities and scaling
functions at IRFP-s\cite{fisheri,fisher}. However, the same type of
asymptotic accuracy of the results is predicted at GFP-s and checked
numerically by the density matrix renormalization group
method\cite{ijl01}.  Therefore we expect the predictions of the SDRG
method about LSFP-s and the SGFP also to be correct. This expectation
finds support in the results of numerical calculations for the $1d$ Heisenberg
model with mixed F-AF couplings\cite{FAF} and for the $\pm J$ square lattice
HAF\cite{2dSG}. Nevertheless alternative calculations are necessary to
check the validity of the predictions of our SDRG results.

Acknowledgment: F.I. is grateful to G. F\'ath for useful
discussions. This work has been supported by a German-Hungarian
exchange program (DAAD-M\"OB), by the Hungarian National Research Fund
under grant No OTKA TO34138, TO37323, MO28418 and M36803, by the
Ministry of Education under grant No. FKFP 87/2001 and by the Centre
of Excellence ICA1-CT-2000-70029.  Numerical calculations are
partially performed on the Cray-T3E at Forschungszentrum J\"ulich.

}


\begin{thebibliography}{99}

\bibitem{patrik}
        P.~Fazekas, {\it Lecture notes on electron correlations and magnetism},
        (World Scientific, Singapore, 1999).

\bibitem{fradkin} E.~Fradkin, {\it Field theories of condensed matter
        systems}, (Addison-Wesley, Redwood City, 1991).


\bibitem{experiment1}
        L.J.~de~Jongh, in {\it Magnetic Phase Transitiona}, ed. M.~Ausloos and
        R.J.~Elliott (Springer, New York, 1983), p. 172.

\bibitem{experiment2}
        D.C.~Johnston, {\it et al.}, Phys. Rev. B {\bf 36}, 4007 (1987);
        S.-W.~Cheong,  {\it et al.}, Phys. Rev. B {\bf 44}, 9739 (1991);
        M.~Corti, A.~Rigamonti, and F.~Tabak, Phys. Rev. B {\bf 52}, 4226 (1995).

\bibitem{MDH}
    S.K.~Ma, C.~Dasgupta, and C.-K.~Hu, Phys. Rev. Lett. 
    {\bf 43}, 1434 (1979);
    C.~Dasgupta and S.K.~Ma, Phys. Rev. B {\bf 22}, 1305 (1980).

\bibitem{fisher}
        D.S.~Fisher, Phys. Rev. B {\bf 50}, 3799 (1994).

\bibitem{ijl01} F.~Igl\'oi, R.~Juh\'asz and P.~Lajk\'o,
        Phys. Rev. Lett. {\bf 86}, 1343 (2001); F.~Igl\'oi,
        Phys. Rev. B {\bf 65}, 064416 (2002).

\bibitem{1dS-1-RG} R.A.~Hyman and K.~Yang, Phys. Rev. Lett. {\bf 78},
        1783 (1997); C.~Monthus, O.~Golinelli, and Th.~Jolicoeur,
        Phys. Rev. Lett. {\bf 79}, 3254 (1997); A.~Saguia, B.~Boechat
        and M.A.~Continentino, Phys. Rev. Lett. {\bf 89}, 117202
        (2002).

\bibitem{1dS-32-RG} G.~Refael, S.~Kehrein, D.S.~Fisher, Phys. Rev. B
        {\bf 66}, 060402 (2002).

\bibitem{ladder_paper} R.~M\'elin, Y.-C.~Lin, P.~Lajk\'o, H.~Rieger
        and F.~Igl\'oi, Phys. Rev.  B {\bf 65}, 104415 (2002).

\bibitem{westerberg} E.~Westerberg, A.~Furusaki, M.~Sigrist and
        P.A.~Lee, Phys. Rev. Lett. {\bf 75}, 4302 (1995); Phys. Rev. B
        {\bf 55}, 12578 (1997).

\bibitem{ladder_dil}
        E.~Yusuf and K.~Yang, preprint-cond-mat/0208458.

\bibitem{kthkmt}
        K.~Kato, {\it et al.} Phys. Rev. Lett. {\bf 84}, 4204 (2000).

\bibitem{sandvik}
        A.W.~Sandvik, Phys. Rev. B {\bf 66}, 024418 (2002).

\bibitem{2dSG-jap}
        Y.~Nonomura and Y.~Ozeki,  J. Phys. Soc. Jpn. {\bf 64}, 2710 (1995). 

\bibitem{2dSG}
        J.~Oitmaa and O.P.~Sushkov, Phys. Rev. Lett. {\bf 87}, 167206
        (2001).

\bibitem{merminwagner}
        N.D.~Mermin and H.~Wagner, Phys. Rev. Lett. {\bf 17}, 1133 (1966).

\bibitem{lutherpeschel}
         A.~Luther and I.~Peschel, \prb {\bf 12}, 3908 (1975).

\bibitem{haldane}
        F.D.M.~Haldane, Phys. Lett. {\bf 93A}, 464 (1983).

\bibitem{ladder}
        For a review, see E.~Dagotto and T.M.~Rice, Science {\bf 271}, 618 (1996).

\bibitem{SZP} H.J.~Schulz, T.A.L.~Ziman and D.~Poilblanc, J. de
        Physique. I {\bf 6}, 675 (1996).

\bibitem{regeryoung}
        J.D.~Reger and A.P.~Young, Phys. Rev. B {\bf 37}, 5978 (1988).

\bibitem{neubergerziman}
        H.~Neuberger and T.A.L.~Ziman, Phys. Rev. B {\bf 39}, 2608 (1989).

\bibitem{triangular} B.~Bernu, P.~Lecheminant, C.~Lhuillier, and
        L.~Pierre, Phys. Rev. B {\bf 50}, 10048 (1994); L.~Capriotti,
        A.E.~Trumper and S.~Sorella, Phys. Rev. Lett.  {\bf 82}, 3899
        (1999); A.E.~Trumper, L.~Capriotti, and S.~Sorella, Phys. Rev.
        B {\bf 61}, 11529 (2000).
        
\bibitem{kagome} C.~Zeng and V.~Elser, Phys. Rev. B {\bf 42}, 8436
        (1990); J.T.~Chalker and J.F.G.~Eastmond, Phys. Rev. B{46},
        14201 (1992); P.W.~Leung and V.~Elser, Phys. Rev. B {\bf 47},
        5459 (1993); P.~Lecheminant, B.~Bernu, C.~Lhuillier, L.~Pierre
        and P.~Sindzingre, Phys. Rev. B {\bf 56}, 2521 (1997);
        C.~Waldtmann, H.-U.~Everts, B.~Bernu, P.~Sindzingre,
        C.~Lhuillier, P.~Lecheminant, and L.~Pierre,
        Eur. Phys. J. B {\bf 2}, 501 (1998).

\bibitem{2dpyr}
        S.E.~Palmer and J.T.~Chalker, preprint cond-mat/0102447.

\bibitem{2dpyr_new}
        R.~M\'elin, unpublished.

\bibitem{pyr}
        B.~Canals and C.~Lacroix, Phys. Rev. Lett. {\bf 80}, 2933 (1998).

\bibitem{J1J2} F.~Figueirido, A.~Karlhede, S.~Kivelson, S.~Sondhi,
        M.~Rocek, and D.S.~Rokhsar, Phys. Rev. B {\bf 41}, 4619 (1989);
        E.~Dagotto and A.~Moreo, Phys. Rev. Lett. {\bf 63}, 2148
        (1989); M.P.~Gelfand, R.R.P.~Singh and D.A.~Huse,
        Phys. Rev. B {\bf 40}, 10801 (1989); R.R.P.~Singh and
        R.~Narayanan, Phys. Rev. Lett. {\bf 65}, 1072 (1990);
        H.J.~Schulz and T.A.L.~Ziman, Europhys.  Lett. {\bf 18}, 355
        (1992).

\bibitem{SOW} M.S.L. du~Croo~de~Jongh, J.M.J.~van~Leeuwen, and 
        W.~van~Saarloos, Phys. Rev.  B {\bf 62}, 14844 (2000); O.P.~Sushkov,
        J.~Oitmaa and Z.~Weihong, Phys. Rev. B {\bf 63}, 104420
        (2001).

\bibitem{MYTT-2ddimer}
        M.~Matsumoto, C.~Yasuda, S.~Todo, and H.~Takayama,
        Phys. Rev. B {\bf 65}, 014407 (2002).

\bibitem{note} In $1d$ where at $\alpha=0$ there is only QLRO the
        system becomes dimerized for any small value of $\alpha$, thus
        $\alpha_c=0$. Here the gap opens with an exponent $\nu=2/3$,
        M.C.~Cross and D.S.~Fisher, Phys. Rev. B {\bf 19}, 402 (1979).

\bibitem{klumper} J.~Sirker, A.~Kl\"umper and K.~Hamacher,
        Phys. Rev. B {\bf 65}, 134409 (2002).

\bibitem{Melin-RG}
        R.~M\'elin, Eur. Phys. J. B {\bf 16}, 261 (2000).

\bibitem{ijr00} F.~Igl\'oi, R.~Juh\'asz, and H.~Rieger, Phys. Rev. B
        {\bf 61}, 11552 (2000).

\bibitem{griffiths}
        R.B.~Griffiths, Phys. Rev. Lett. {\bf 23}, 17 (1969).

\bibitem{phenom} M.J.~Thill and D.A.~Huse, Physica (Amsterdam) {\bf
        15A}, 321 (1995); H.~Rieger and A.P.~Young, Phys. Rev. B {\bf
        54}, 3328 (1996); F.~Igl\'oi, R.~Juh\'asz and H.~Rieger,
        Phys. Rev. B {\bf 59}, 11308 (1999).

\bibitem{YTMT02} C.~Yasuda, S.~Todo, M.~Matsumoto and H.~Takayama,
        preprint-condmat/0204397.

\bibitem{mila_kagome}
        F.~Mila, preprint-cond-mat/9805078.

\bibitem{sg-review}
        For a review see e.g.: H.~Rieger and A.P.~Young,
        {\it Quantum Spin Glasses}, Lecture Notes in Physics {\bf 492},
        in ``Complex Behaviour of Glassy Systems'', p.\ 254,
        ed.\ J.M.~Rubi and C.~Perez-Vicente 
        (Springer Verlag, Berlin-Heidelberg-New$\,$York, 1997).
  
\bibitem{fisheri}
        D.S.~Fisher, Phys. Rev. Lett. {\bf 69}, 534 (1992); 
        Phys. Rev. B {\bf 51}, 6411 (1995).

\bibitem{2drg} O.~Motrunich, S.-C.~Mau, D.A.~Huse and D.S.~Fisher,
        Phys. Rev. B {\bf 61}, 1160 (2000).

\bibitem{lkir} Y.-C.~Lin, N.~Kawashima, F.~Igl\'oi and H.~Rieger,
        Prog. Th. Phys. (Suppl.) {\bf 138}, 470 (2000); D.~Karevski,
        Y.-C.~Lin, H.~Rieger, N.~Kawashima and F.~Igl\'oi,
        Eur. Phys. J. B {\bf 20}, 267 (2001).

\bibitem{note2} The numerical results obtained in Ref.~\onlinecite{2dSG-jap}
        suggest that weakly frustreted systems, such as the randomly
        frustrated models in Sec. IV.4 with a small fraction of F
        bonds, are still in the random HAF phase and not in the SG phase.

\bibitem{boseglass} A similar relation, $\omega=0$, and thus $z=d$,
        holds at the boson localization transition point:
        M.P.A.~Fisher, P.B.~Weichman, G.~Grinstein and D.S.~Fisher,
        Phys. Rev. B {\bf 40}, 546 (1989); and in the Bose glass phase:
        J.~Kisker and H.~Rieger, Physica A {\bf 246}, 348 (1997). This
        singularity is controlled by a GFP, therefore its origin is
        different from that at the SGFP.

\bibitem{FAF} T.~Hikihara, S.~Furusaki, and M.~Sigrist,
        Phys. Rev. B {\bf 60}, 12116 (1999).

\end{thebibliography}
\end{document}